\documentclass[twocolumn,aps,prl,english,floatfix,superscriptaddress]{revtex4}
\usepackage{amsmath,amsfonts,amssymb,graphicx,bm,color,mathrsfs,verbatim}
\pdfoutput=1

\usepackage{soul}
\usepackage{ulem}
\usepackage{cancel}

\newcommand{\be}{\begin{equation}}
\newcommand{\ee}{\end{equation}}
\newcommand{\beq}{\begin{eqnarray}}
\newcommand{\eeq}{\end{eqnarray}}
\newcommand{\bml}{\begin{multline}}
\newcommand{\eml}{\end{multline}}

\begin{document}

\title{Comment on ``Floquet Fractional Chern Insulators''}
\author{Luca D'Alessio}
\affiliation{Department of Physics, The Pennsylvania State University, 
University Park, PA 16802, USA}
\affiliation{Department of Physics, Boston University, Boston, MA 02215, USA}

\begin{abstract}
In a recent paper~\cite{Adolfo}, Grushin et al. studied a Fermi-Hubbard-model
for spinful electrons on a honeycomb lattice coupled to an external
polarized electric field. By computing the Floquet Hamiltonian perturbatively
to order $\omega^{-1}$ (where $\omega$ is the frequency of the drive)
they predicted that, for specific values of the driving and filling,
the system can transition into a Fractional Chern Insulator. In this
comment we point out that: i) the calculation
of the Floquet Hamiltonian misses some terms of order $\omega^{-1}$ and ii) the assumption that
the Floquet bands are filled as in time-independent systems is questionable. 
\end{abstract}
\maketitle

In Ref.~\cite{Adolfo}, Grushin et al. considered a Fermi-Hubbard model
for spinful electrons on an honeycomb lattice coupled to a circularly
polarized electric field, $\vec{E}(\tau)=E_{0}\left(\cos(\omega\tau),-\sin(\omega\tau)\right)$.
(We use $\tau$ to represent time since we want to reserve $t_{1}$
for the nearest-neighbor hopping.) The electromagnetic gauge is fixed
by choosing to represent the electric field via a vector potential
$\vec{A}(\tau)=A_{0}\left(\sin(\omega\tau),\cos(\omega\tau)\right)$
which induces a time-dependent phase in the hopping. From the relation
$\vec{E}=\frac{\partial\vec{A}}{\partial\tau}$ it is easy to see
that the amplitude of the electric field and the vector potential are
related by $E_{0}=\omega A_{0}$. The total Hamiltonian is $H(t)=H_{0}(\tau)+H_{int}$
where: 
\beq
H_{0}(\tau)&\equiv-t_{1}\sum_{\langle i,j\rangle}\sum_{\sigma}\left(e^{-i\varphi_{i,j}(\tau)}c_{i,\sigma}^{\dagger}c_{j,\sigma}+h.c\right) \nonumber\\
H_{int}&\equiv U\,\sum_{i}\, n_{i,\uparrow}n_{i,\downarrow}+V\sum_{\langle i,j\rangle}\sum_{\sigma,\sigma'}n_{i,\sigma}n_{j,\sigma'} \nonumber
\eeq
and $\langle i,j\rangle$ are nearest neighbor 
lattice sites, $\varphi_{i,j}(\tau)=\frac{e}{\hbar}\vec{A}(\tau)\cdot\left(\vec{r_{i}}-\vec{r}_{j}\right)$ 
is the time-dependent phase induced by the vector potential and $n_{i,\sigma}=c_{i,\sigma}^{\dagger}c_{i,\sigma}$
is the spin-dependent electron density at site $i$. The interaction
parameters are $U=3t_{1}$ and $V=2t_{1}$.

The total Hamiltonian is time-periodic with period $T=\frac{2\pi}{\omega}$
and the evolution operator over a period defines the Floquet Hamiltonian
via the relation $U(T,0)=e^{-iH_{F}T/\hbar}$. In the high frequency
limit the Floquet Hamiltonian can be computed systematically using
the Magnus Expansion~\cite{Magnus} (ME), i.e. $H_{F}=\sum_{n}\, H_{F}^{(n)}$. In
the chosen electromagnetic gauge (see below for details) the $n$-th
term in the ME is of order $\omega^{-n}$ and therefore the Floquet
Hamiltonian to order $\omega^{-1}$ is given by $H_{F}\approx H_{F}^{(0)}+H_{F}^{(1)}+\mathcal{O}\left(\omega^{-2}\right)$.
The zero order term in this series is the time-average Hamiltonian
while the $\omega^{-1}$ correction is given by~\cite{Magnus}:
\begin{equation}
H_{F}^{(1)}=\frac{1}{2(i\hbar)T}\int_{0}^{T}d\tau_{1}\int_{0}^{\tau_{1}}d\tau_{2}\,\left[H(\tau_{1}),H(\tau_{2})\right]\label{eq:correct}
\end{equation}
The oversight in Ref.~\cite{Adolfo} is that the authors replace,
in the expression for $H_{F}^{(1)}$, the total Hamiltonian $H(\tau)$
with the single particle Hamiltonian $H_{0}(\tau)$ effectively calculating:
\begin{equation}
\widetilde{H}_{F}^{(1)}=\frac{1}{2(i\hbar)T}\int_{0}^{T}d\tau_{1}\int_{0}^{\tau_{1}}d\tau_{2}\,\left[H_{0}(\tau_{1}),H_{0}(\tau_{2})\right]\label{eq:mistake}
\end{equation}
The Hamiltonians $\widetilde{H}_{F}^{(1)}$ and $H_{F}^{(1)}$ differ
by terms of order $\omega^{-1}$ potentially invalidating the main result in Ref.~\cite{Adolfo}.
The Floquet Hamiltonian to order $\omega^{-1}$
has been worked out explicitly in Ref.~\cite{Marin} and contains,
beside the terms analyzed in Ref.~\cite{Adolfo}, interaction assisted
tunneling. Schematically:
\[
H_{F}^{(1)}=\widetilde{H}_{F}^{(1)}+\frac{1}{\omega}\,(\text{interaction assisted tunneling})
\]

This oversight in Ref.~\cite{Adolfo} is perhaps due to a confusion
between different electromagnetic gauges. In fact the electric field
can be described via a scalar potential so that the single particle Hamiltonian is: 
\[
H_{0}^{\prime}(\tau)\equiv-t_{1}\sum_{\langle i,j\rangle}\sum_{\sigma}\left(c_{i,\sigma}^{\dagger}c_{j,\sigma}+h.c\right)+\sum_{i,\sigma}\Phi(i,\tau)\, n_{i,\sigma}
\]
where $\Phi(i,\tau)=-e\,\vec{E}(\tau)\cdot\vec{r}_{i}$ is the (time-periodic)
scalar potential which couples to the density. Therefore \textit{in this electromagnetic
gauge} the driving commutes with the interaction and, in the expression for $H_{F}^{(1)}$, the total Hamiltonian can be replaced by the single particle Hamiltonian. This seems
to lead to the unphysical conclusion that the presence of interaction
assisted tunneling in the Floquet Hamiltonian to order $\omega^{-1}$
depends on the electromagnetic gauge in which the calculation is performed.
The resolution of this contradiction is that, in this second electromagnetic
gauge, the ME is \textit{not} an expansion in $\omega^{-1}$ and to obtain 
the Floquet Hamiltonian to order $\omega^{-1}$ the entire ME needs to be resummed~\cite{Marin}.  
This can be understood by noticing that $E_{0}=\omega A_{0}$ and therefore
the amplitude of the electric field diverges in the high-frequency
limit (we are assuming, as in Ref.~\cite{Adolfo}, $A_0=\text{const.}$ and $\omega\rightarrow\infty$)
making this second electromagnetic gauge very inconvenient to compute $H_F$ perturbatively.
When the ME is resummed the interaction assisted tunneling is correctly reproduced.

Finally in Ref.~\cite{Adolfo} the authors claim that,
since the Floquet Hamiltonian is time-independent its bands can be
filled ``as in the case of time-independent systems''. Here the
authors seem to assume that when Floquet systems are connected to
a heat bath the occupation of the Floquet states is thermal with the
temperature of the bath. This is not generally correct and the occupation of the Floquet states
might be non-thermal (see for example~\cite{non-thermal})
even in the limit of high-frequency driving.
Moreover, in absence of the bath, interacting periodically driven systems are expected to heat up towards infinite temperature~\cite{luca}.

\end{document}